\documentclass[review]{elsarticle}
\usepackage{amsmath,amssymb,graphicx}
\usepackage{epstopdf}
\usepackage{grffile}
\usepackage{booktabs}
\usepackage{rotating}
\usepackage{subcaption}
\usepackage{multirow}
\usepackage[pdftex,colorlinks=false]{hyperref}
\usepackage{textcomp}
\journal{Acta Astronautica}
\begin{document}

\begin{frontmatter}

\title{A four-pole power-combiner design for far-infrared and submillimeter spectroscopy}

\author[NASA,MIT]{Giuseppe Cataldo\corref{cor1}}
\ead{Giuseppe.Cataldo@NASA.gov, gcataldo@mit.edu}
\author[NASA]{Samuel H. Moseley}
\author[NASA]{Edward J. Wollack}
\cortext[cor1]{Corresponding author}
\address[NASA]{NASA Goddard Space Flight Center, United States of America}
\address[MIT]{Massachusetts Institute of Technology, United States of America}

\begin{abstract}
The far-infrared and submillimeter portions of the electromagnetic spectrum provide a unique view of the astrophysical processes present in the early universe. Micro-Spec ($\mu$-Spec), a high-efficiency direct-detection spectrometer concept working in the 450--1000-\textmu m wavelength range, will enable a wide range of spaceflight missions that would otherwise be challenging due to the large size of current instruments and the required spectral resolution and sensitivity.
This paper focuses on the $\mu$-Spec two-dimensional multimode region, where the light of different wavelengths diffracts and converges onto a set of detectors. A two-step optimization process is used to generate geometrical configurations given specific requirements on spectrometer size, operating spectral range, and performance. The canonically employed focal-plane constraints for the power combiner were removed to probe the design space in its entirety. A new four-stigmatic-point optical design solution is identified and explored for use in far-infrared and submillimeter spectroscopy.
\end{abstract}

\end{frontmatter}


\section{Introduction}
\label{sec:introduction}
Far-infrared (IR) and submillimeter (15~\textmu m to 1~mm) spectroscopy provides a powerful tool to probe a wide range of environments in the universe. In the past thirty years, discoveries made by several space-based observatories have provided unique insights into physical processes leading to the evolution of the universe and its contents.
This information is encoded in a variety of molecular and fine structure lines; observations of such spectral lines enable the exploration of galaxies at high redshifts. The fine structure lines of abundant elements (carbon, nitrogen, and oxygen), for example, allow tracing the obscured star formation and Active Galactic Nuclei (AGN) activity into the high-redshift universe. One can measure galaxy redshifts and determine their elemental abundances and physical conditions out to redshifts $z>5$.
In spite of this, a number of questions remain unanswered regarding the very early steps of the universe as well as galactic, stellar, and planetary formation. The ability to explore this rich spectral region has been limited by the size and cost of the cryogenic spectrometers required to carry out these measurements.
The work proposed here specifically addresses the need for integrated spectrometers and background-limited far-IR direct detectors. For space-borne astrophysics systems, the specific requirements are shown in Table~\ref{tab:AstrophysicsNeeds} and compared against the current state of the art~\cite{NASA-STR}.
\begin{table}[!h]
	\center
	\caption{Summary of far-IR cryogenic spectrometer and detector array requirements and comparison with current state of the art~\cite{NASA-STR}.}
	\label{tab:AstrophysicsNeeds}
	\begin{tabular}{lll}
		\toprule
		 \textbf{Metric} & \textbf{State of the art} & \textbf{Requirements} \\
		\midrule
		 Wavelength, $\lambda$							& $250-700$~\textmu m                 & $220-2000$ \textmu m \\
		 Noise Equivalent Power, NEP				& $10^{-19}$ W/$\sqrt{\mbox{Hz}}$ & $<10^{-20}$ W/$\sqrt{\mbox{Hz}}$ \\
		 Spectral resolution, $\cal R$								& $\geq 100$			  							& $\geq 1200$ \\
		 Detective Quantum Efficiency, DQE	& $\sim15\%$                      & $> 90\%$ \\
		 Time constant, $\tau$							& 100 ms													& $< 10$ ms \\
		\bottomrule
	\end{tabular}
\end{table}

In order to realize the goals outlined in Table~\ref{tab:AstrophysicsNeeds}, a high-performance integrated spectrometer module, Micro-Spec ($\mu$-Spec), operating in the 450--1000-\textmu m (300--650-GHz) range is proposed.
$\mu$-Spec can be compared to a grating spectrometer~\cite{Rowland,Yen, Marz, WuChen, Munoz}, in which a plane wave is reflected from the grating and the phase of each partial wave scattered from the rulings is a linear function of position across the grating. An example of planar Rowland grating architecture is Z-Spec, in which propagation occurs in parallel-plate waveguides~\cite{Naylor, Bradford2003, Bradford2004, Earle}. Another comparison can be made with one-dimensional bootlace lenses found in microwave practice~\cite{Rotman, Katagi, Hansen, Rappaport}, which $\mu$-Spec builds upon for submillimeter wave applications. Finally, another variation used at millimeter wavelengths, which does not rely on optical interference as in grating spectrometers, is a narrow-band filter-bank spectrometer. Examples realized in superconducting transmission lines are SuperSpec~\cite{Kovacs, Shirokoff, Hailey-Dunsheath} and the Delft SRON High-Z Mapper (DESHIMA)~\cite{DESHIMA}.

\begin{figure*}[!htbp]
	\centering
	\vspace{24pt}
		\includegraphics[width=1.00\textwidth]{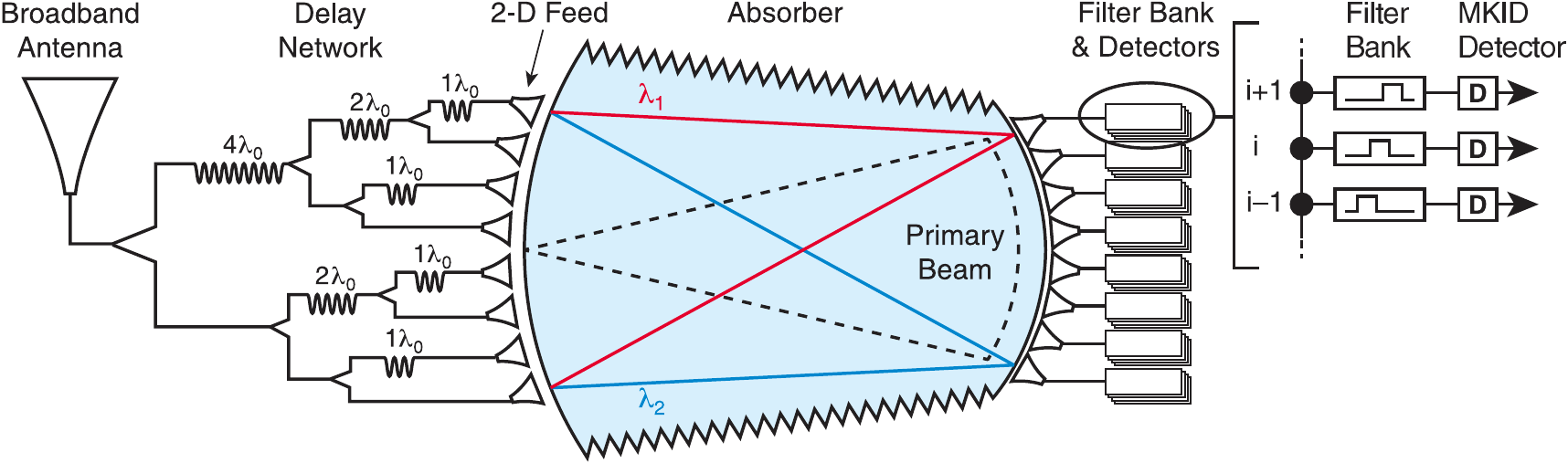}
			\caption{Layout of an individual $\mu$-Spec wafer. The radiation is coupled into the instrument through a broadband antenna and is then transmitted through a superconducting transmission line to a divider and a phase delay network. The spectrum enters the multimode region through an array of feeds which concentrates the power along the focal surface with different wavelengths at different locations. The receivers are connected to a bank of order-sorting filters and MKID detectors~\cite{Cataldo2014}. Multiple spectrometer wafers can be packaged and potentially used in defining an instrument system.}
	\label{fig:Instrument-layout-F1}
\end{figure*}

$\mu$-Spec differs from these approaches by the order of processing of the light in the spectrometer. In $\mu$-Spec (Fig.~\ref{fig:Instrument-layout-F1}), the incoming radiation collected by the telescope is coupled to the spectrometer via a broadband dual-slot antenna used in conjunction with a hyperhemispherical silicon lens and directed to a series of power splitters and a delay network made of superconducting microstrip transmission lines. Analogous to the Rowland grating~\cite{Rowland}, the delay network creates a phase retardation across the input to a planar-waveguide multimode region, which has two internal planar antenna arrays, one for transmitting and one for receiving the radiation as a function of wavelength. Absorber structures lining the multimode region terminate the power emitted into large angles or reflected from the receiver antenna array. An array of planar feed structures is employed to couple the radiation to the multimode region and concentrates the power along the focal surface with different wavelengths at different locations. The outputs are connected to a bank of order-sorting filters which terminate the power in an array of microwave kinetic inductance detectors (MKIDs) for detection and read-out.
The entire spectrometer circuit is integrated on a $\sim$~10-cm$^2$ silicon chip (i.e., the hyperhemispherical lens, relay optics, and telescope are not on the chip and are part of the instrument system). This compact footprint is accomplished through the use of single-mode microstrip delay lines, which can compactly be folded on the silicon wafer and reduce the required physical line length by a factor of the medium's effective refractive index.

The frequency range of the implementation presented here is limited to wavelengths $\lambda > 250$~\textmu m by the gap frequency of currently available low-loss superconductors. These include niobium (Nb) and niobium-titanium nitride (NbTiN) for the transmission line structures, and molybdenum nitride (MoN) for the detectors.
This paper will describe the design process of the $\mu$-Spec multimode region and illustrate the results in terms of geometry, imaging quality, and efficiency.

\section{Multimode region design}
\label{sec:design}
In designing a spectrometer, it is possible to define points on the focal plane where the phase error of the diffracted light is identically equal to zero. These points are called stigmatic points.
Increasing the number of such points on the focal surface presents the advantage of improving a spectrometer's imaging quality, which results in lowering the overall phase error on the entire focal plane and increasing the usable spectral bandwidth. As a consequence, the number of spectrometer channels and the resolving power can grow.

Examples of designs with two stigmatic points can be found in the literature~\cite{Marz, Munoz, Naylor}. We built upon these designs to generate spectrometer concepts with an increased number of stigmatic points.
A three-stigmatic-point prototype version with a resolving power $\cal R$ $=65$ in first order $(M=1)$ was designed~\cite{Cataldo2014} and built, and is currently under evaluation at the NASA Goddard Space Flight Center. Additional designs are described in~\cite{Cataldo2014b} for configurations with resolving powers equal to $\cal R$~$=260$ and $\cal R$~$=520$ in higher order ($M>1$). These designs were obtained through a constrained optimization process in which zero phase error was imposed on three preselected points. However, a fourth stigmatic point was observed beyond the angular range in use. It is the purpose of the work presented in this paper to show how to use this additional degree of freedom to increase the number of spectrometer channels and resolving power. In the following section, therefore, we describe a design for $\cal R$~$\approx260$ in first order with four stigmatic points, as a result of an unconstrained optimization process which maximizes the instrument resolving power and minimizes the root-mean-square (RMS) phase error on the focal plane.

\subsection{Problem formulation}
\label{sec:formulation}
\begin{figure}[!ht]
	\centering
	\includegraphics[width=0.80\textwidth]{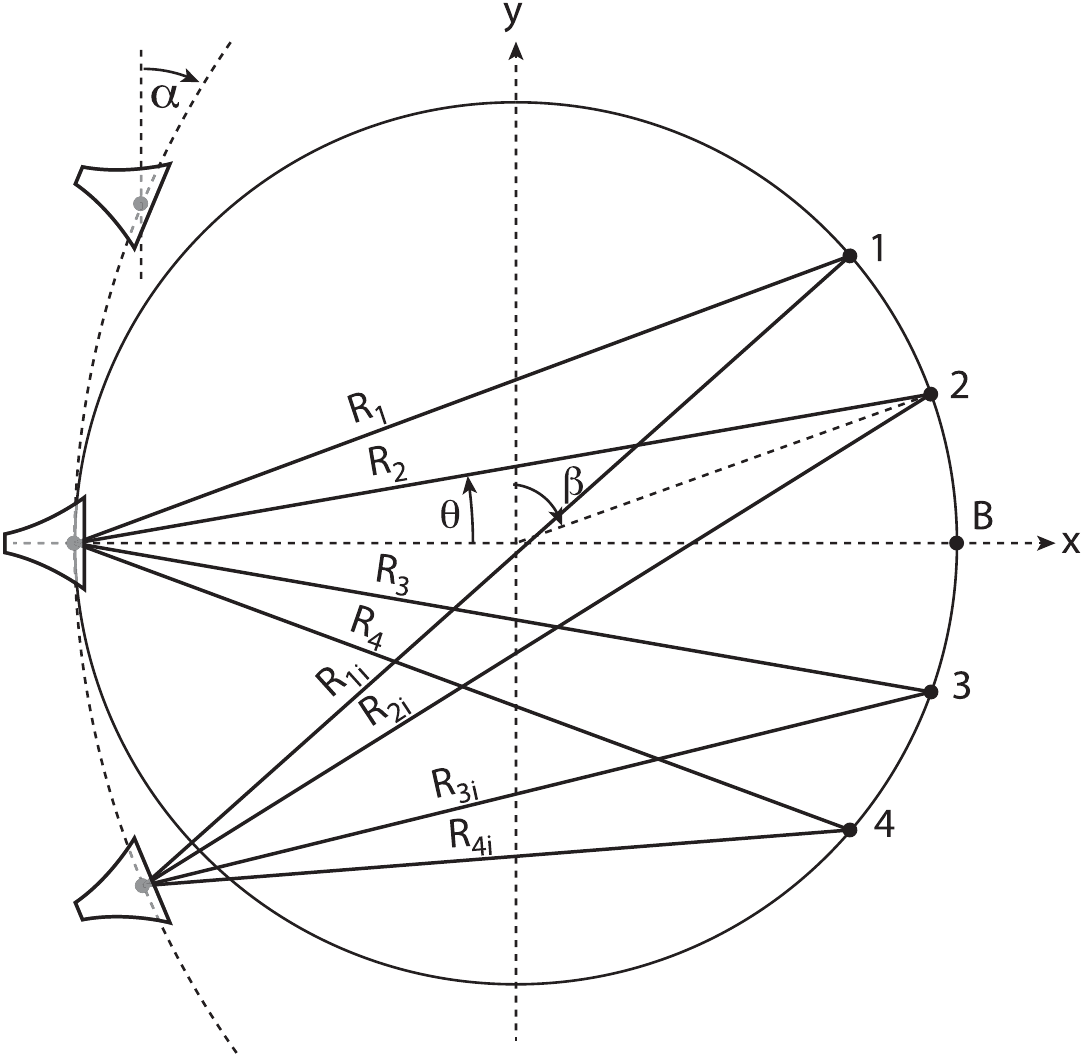}
	\caption{Simplified representation of the grating geometry. On the left side three radiators can be seen, which point to the blaze point, B. The solid lines represent the paths to four stigmatic points from a generic $i$-th radiator's phase center ($R_{1i},R_{2i},\ldots$) and the central reference feed ($R_{1},R_{2},\ldots$).}
	\label{fig:InstrumentSimplifiedVersion}
\end{figure}

As explained in~\cite{Cataldo2014}, the design variables are the $x$ and $y$ coordinates of the $N_e$ emitters' centers and the electrical path lengths in silicon, $R_i^e$, for each feed's electrical delay (Fig.~\ref{fig:InstrumentSimplifiedVersion}).
The resolving power is defined as
\begin{equation}
	{\cal R} = M\cdot N_e,
	\label{eq:R}
\end{equation}
where $M$ is the order of the spectrometer and $N_e$ is a power of 2, given the structure of the power divider network (Fig.~\ref{fig:Instrument-layout-F1}).

The first step of the design consists of finding the maximum achievable resolving power, ${\cal R}_{max}$, as a function of $M$ and the relative emitter pitch, $\eta = p/\lambda_{avg}$ ($p =$ emitter pitch, $\lambda_{avg} =$ central wavelength associated with the geometric average frequency), given specific requirements on spectrometer radius, $R$, and operating spectral range, as well as certain constraints on performance.

\noindent The formulation of this mixed integer non-linear problem is as follows:
\begin{eqnarray}
	\label{eq:problem}
	\max {\cal R}_{max}(M,\eta) &=& M/\eta \cdot R/\lambda_{avg}\\
	\label{eq:He}
	\mbox{subject to } \qquad H_e(M,\eta)&\leq&R \\
	\label{eq:Hr}
	H_r(M,\eta)&\leq&\pi R \\
	\label{eq:Rmax}
	{\cal R}_{max}(M,\eta)&>&200 \\
	\eta&>&0\\
	\label{eq:M}
	M &\geq& 1,\quad M \mbox{ integer}
\end{eqnarray}
Eq.~\eqref{eq:He} imposes that the width of the emitter array, $H_e$, be less than or equal to the spectrometer radius so that there is no aberration; Eq.~\eqref{eq:Hr} lets the receiver array be as large as the focal plane to maximize its utilization; and Eq.~\eqref{eq:Rmax} sets a minimum value for the required maximum resolving power, ${\cal R}_{max}$, thereby eliminating solutions in which we are not interested. Finally we note that, to enlarge the tradespace, the constraint of equal emitter and receiver pitch, used in our previous designs~\cite{Cataldo2014,Cataldo2014b}, was removed.
The entire problem was solved with a Branch and Bound algorithm~\cite{NehmauserWolsey} using the Interior Point OPTimizer (IPOPT)~\cite{IPOPT} and Coin-or Branch and Cut (CBC)~\cite{CBC} solvers.

Table~\ref{tab:optimalPar} shows the requirements on spectrometer size and spectral range used for this problem. The minimum and maximum frequencies are no longer associated with any stigmatic point and they fall within the spectral range defined in Table~\ref{tab:AstrophysicsNeeds}. The average frequency was computed as their geometric mean.

\begin{figure*}[htbp]
	\centering
	\begin{subfigure}[b]{.47\textwidth}
		\includegraphics[width=1.00\textwidth]{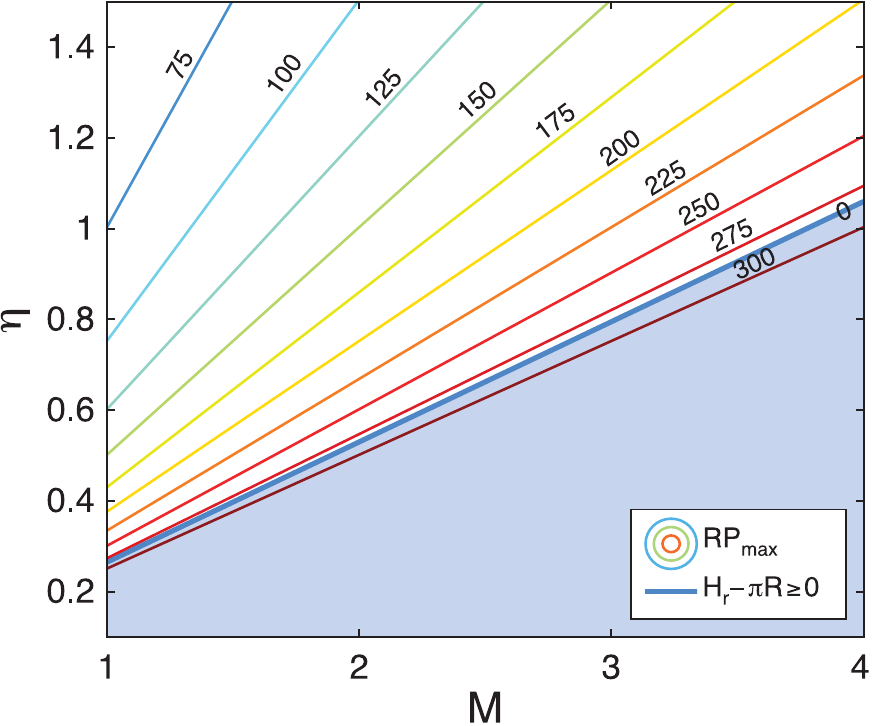}
	\caption{The contour plot depicts all the feasible values of ${\cal R}_{max}$ associated with different values of $M$ and $\eta$. For $M=1$ and $\eta=0.2916$, the optimal solution is ${\cal R}_{max} = 275$.}
	\label{fig:Tradespace_R250max_M1}
	\end{subfigure}
	\qquad
	\begin{subfigure}[b]{.47\textwidth}
		\includegraphics[width=1.00\textwidth]{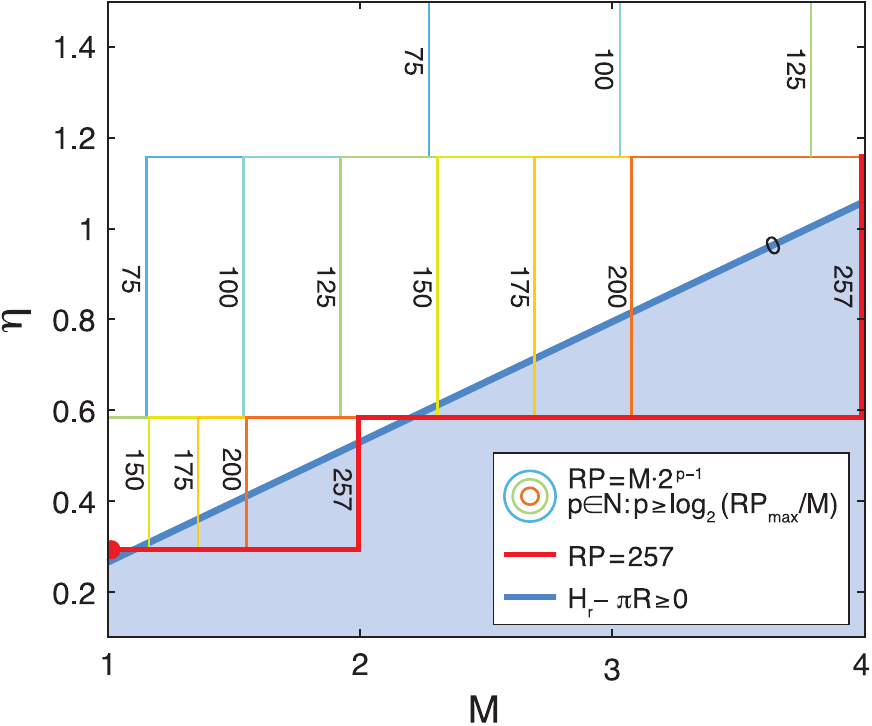}
		\caption{The contour plot represents the feasible values of ${\cal R}$ for all orders given as powers of 2. The optimal solution for $M=1$ and $\eta=0.2916$ is associated with a resolving power ${\cal R} =257$.}
		\label{fig:Tradespace_R250_M1}
	\end{subfigure}
	\caption{Objective spaces of Problem~\eqref{eq:problem}. The blue area represents the infeasible region corresponding to the active constraint in Eq.~\eqref{eq:Hr}. Both plots show that, for each order $M$, several feasible solutions exist for different values of $\eta$.}
	\label{fig:R250M1}
\end{figure*}

The objective spaces as a function of $M$ and $\eta$ are shown in Fig.~\ref{fig:R250M1}. On the left (Fig.~\ref{fig:Tradespace_R250max_M1}), it is possible to visualize the feasible objective space of the optimization problem described above for ${\cal R}_{max}$ along with the active constraint, Eq.~\eqref{eq:Hr} (blue area).
The feasible solutions of Eq.~\eqref{eq:problem} populate that part of the contour plot above the blue area, whereas the optimal solutions lie at the intersection of the active constraint with the largest ${\cal R}_{max} \approx 275$, one for each value of $M$.
One can thus choose the highest desired spectrometer order, $M_{max}$, which is associated with the highest design frequency band. The frequency bands corresponding to orders $M<M_{max}$ can be calculated by scaling the highest one by a factor of $M/M_{max}$.
We decided to investigate the first-order case as an example of simple and robust system. Higher-resolution instruments will certainly require higher-order operations. Table~\ref{tab:optimalRP} shows the values of the design variables associated with this particular optimal solution as well as the values of Eqs.~\eqref{eq:He}-\eqref{eq:M}, which satisfy these constraints.

After solving this problem, it was possible to compute the number of emitters, $N_e={\cal R}_{max} /M$, and round it down to a power of 2. According to Eq.~\eqref{eq:R}, this causes the actual resolving power, ${\cal R}$, calculated with this updated value of $N_e$, to be lower than ${\cal R}_{max}$.
The plot in Fig.~\ref{fig:Tradespace_R250_M1} shows the values of ${\cal R}$ (red line) and the values of $M$ and $\eta$ that would make such realizations possible. In particular, for a first-order design ($M=1$), the optimal solution corresponds to a resolving power ${\cal R}=257$ with a relative emitter diameter $\eta=0.2916$.
\begin{table}[!ht]
	\center
	\caption{Requirements on spectrometer size and spectral range.}
	\label{tab:optimalPar}
	\begin{tabular}{llll}
		\toprule
		\textbf{Parameter} & \textbf{Symbol} & \textbf{Unit} & \textbf{Value} \\
		\midrule
			Multimode region radius & $R$		& cm    & 1.25  \\
			Minimum frequency & $f_{min}$   & GHz   & 430.0 \\
			Maximum frequency & $f_{max}$   & GHz   & 650.0 \\
			Average frequency & $f_{avg}$   & GHz   & 528.7 \\
		\bottomrule
	\end{tabular}
\end{table}
\begin{table}[!ht]
	\center
	\caption{Optimal solution of Eq.~\eqref{eq:problem} for $M=1$.}
	\label{tab:optimalRP}
	\begin{tabular}{llll}
		\toprule
		\textbf{Variable}  					& \textbf{Symbol}	 & \textbf{Unit} & \textbf{Value} \\
		\midrule
			Spectrometer order				& $M$ 		         & -      & 1        	\\
			Relative emitter diameter & $\eta$           & -      & 0.2916   	\\
			Maximum resolving power		& ${\cal R}_{max}$ & -      & 275   		\\\hline
			Emitter array width	  		& $H_e$            & cm     & 1.2452   	\\
			Receiver array width	 		& $H_r$            & cm     & 3.5674   	\\
			Emitter pitch							& $p$   	         & cm     & 0.0048   	\\
			Receiver pitch 						& $s$ 	           & cm     & 0.0167   	\\\hline
			Number of emitters	 			& $N_e$            & -      & 257       \\
			Number of receivers 			& $N_r$            & -      & 149       \\
			Resolving power 					& ${\cal R}$       & -      & 257 \\
		\bottomrule
	\end{tabular}
\end{table}

The second step toward determining the optimal solution in terms of the above-mentioned design variables consists of minimizing the overall RMS phase error, $\varphi_{\mbox{\scriptsize{RMS}}}$, through a figure of merit representing the area subtended by $\varphi_{\mbox{\scriptsize{RMS}}}$ on the focal plane as follows:
\begin{equation}
	\label{eq:RMS}
	\min \int_0^\pi {\varphi_{\mbox{\scriptsize{RMS}}}(\beta)\;\mathrm{d}\beta},
\end{equation}
with
\begin{equation}
	\label{eq:RMS2}
 	\varphi_{\mbox{\scriptsize{RMS}}} = \sqrt{\sum_{i=1}^{N_e} \frac{[\varphi_{i}(x_i,y_i,R^e_i,\beta)-\left<\varphi(\beta)\right>]^2}{N_e}}.
\end{equation}
Here, $\varphi_{i}$ is the relative phase of each transmitter, $\left< \varphi(\beta) \right> = 0$ is the relative phase of the central transmitter (this is zero by construction as the central radiator is used as a reference) and $\beta$ represents the angle corresponding to each of the points in which the focal plane was discretized (Fig.~\ref{fig:InstrumentSimplifiedVersion}). When setting Eq.~\eqref{eq:RMS2} equal to zero, its analytical expression is a fourth-order periodic function of $\sin\beta$ and $\cos\beta$ with coefficients depending on $x_i$, $y_i$, and $R^e_i$ $(i=1,\ldots, N_e)$. The apodization of the feed illumination limits the domain of interest to $0\leq\beta\leq\pi$. In numerically exploring this function over this range, a maximum of four real roots could be identified, which repeat themselves with a periodicity of $2\pi$.

\subsection{Optimization results}
\begin{figure}[htbp]
	\centering
			\includegraphics[width=0.80\textwidth]{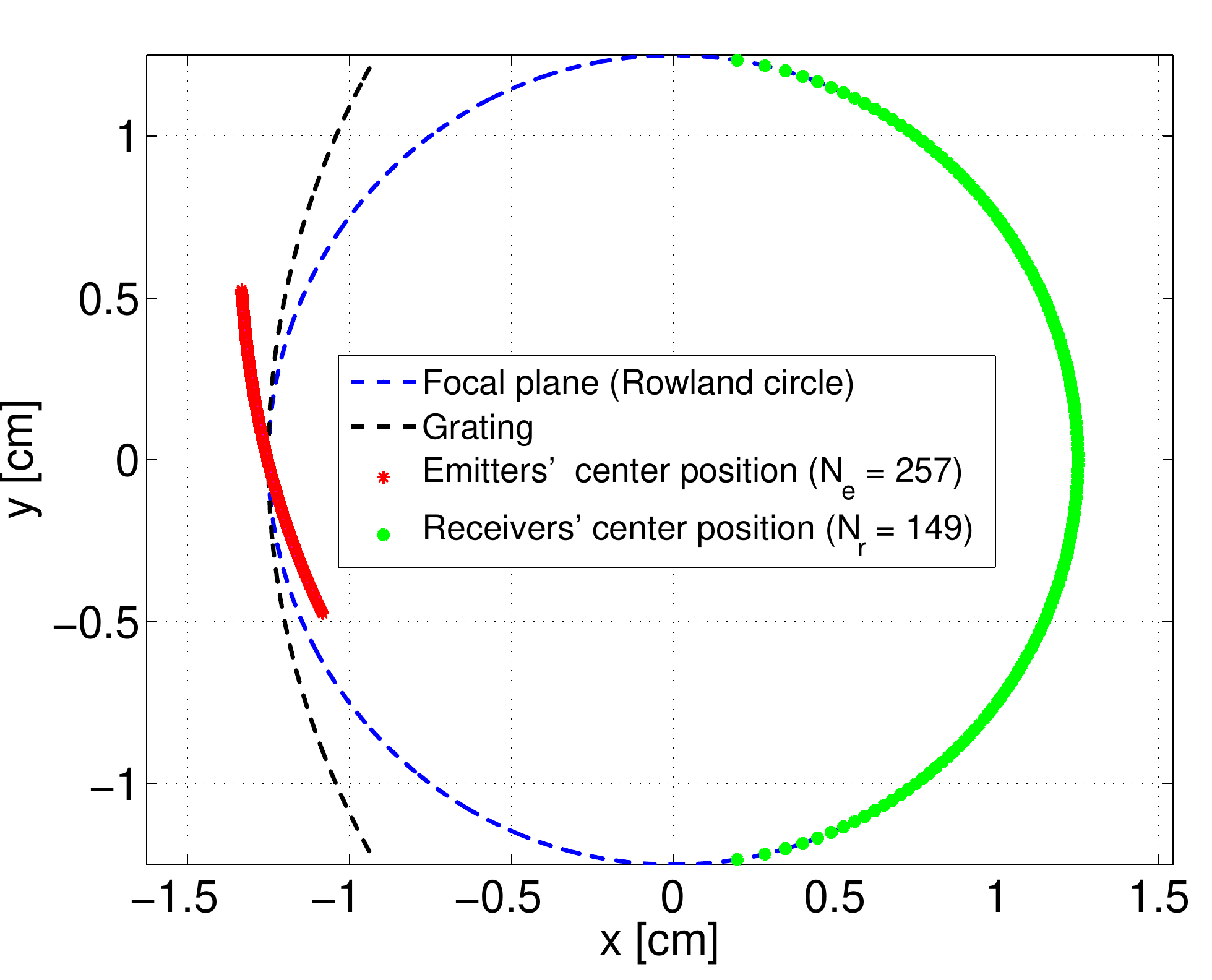}
	\caption{Optimized multimode region design with a resolving power ${\cal R}=257$ and order $M=1$.}
	\label{fig:Micro-Spec Final Design RP257}
\end{figure}

The solution to the minimization problem defined in Eq.~\eqref{eq:RMS} was found with a quasi-Newton algorithm~\cite{Fletcher} and can be seen in Fig.~\ref{fig:Micro-Spec Final Design RP257}. The emitters' positions are indicated in red and present several characteristics similar to those found and discussed in~\cite{Cataldo2014b}. First, they do not lie on the grating circle but on a curve that is tilted leftwards and intersects the grating circle at the central emitter before ending up inside the multimode region. Second, it was verified that the shape of this curve only approximates a circle with a radius $\sim 2.2\,R$ and is not symmetric. In the case presented here, this is caused by the absence of constraints on all stigmatic points. In general, for a two-stigmatic-point configuration the emitters lie exactly on the grating circle, as previously reported in the literature~\cite{Rowland,Naylor}. The imposition of zero RMS phase error on a third stigmatic point (the blaze point) also caused a similar tilting effect~\cite{Cataldo2014b}, unless the emitters could be constrained to lie within a small distance (e.g., $\lambda_1/8$) from the $2R$ circle~\cite{Cataldo2014}.

\begin{figure}[!ht]
	\centering
		\includegraphics[width=0.95\textwidth]{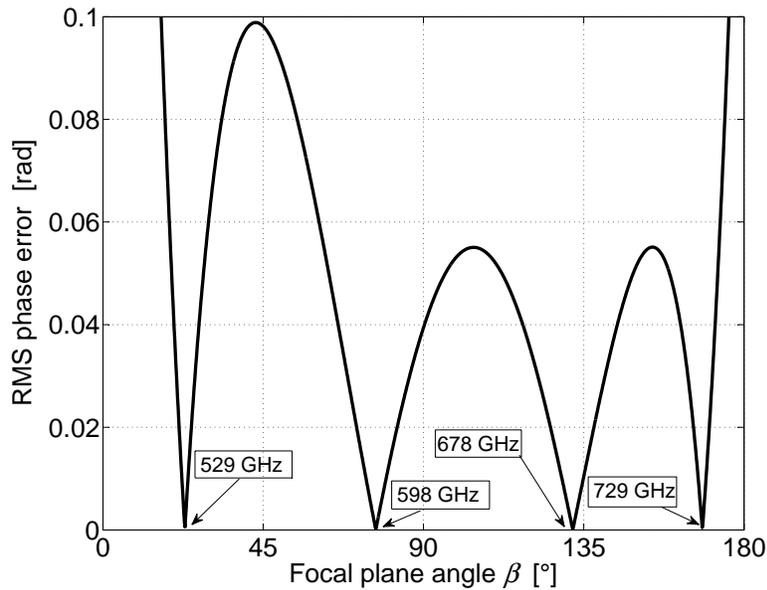}
	\caption{RMS phase error distribution on the focal plane. The worst peak value is 0.1~rad at almost 45$^\circ$, and four stigmatic points can be seen over a 160$^\circ$ angular range.}
	\label{fig:RMS Phase Error}
\end{figure}

The RMS phase error is shown in Fig.~\ref{fig:RMS Phase Error}. Its values remain below 0.1~rad over an angular range spanning approximately from 16$^\circ$ to 176$^\circ$. This phase error does not lead to a significant defocus of the light in the spectrometer~\cite{Ruze} and represents a $\sim30\%$ improvement in the focal plane utilization over the previous designs~\cite{Cataldo2014,Cataldo2014b}.
Four stigmatic points are visible, but they are no longer associated with a predefined frequency, given the absence of constraints on them. In Fig.~\ref{fig:RMS Phase Error} it can be seen to what frequencies they correspond in this design. The nominal spectral range indicated in Table~\ref{tab:optimalPar} is only partially covered down to $\sim$ 510~GHz at 0$^\circ$, while above 130$^\circ$ frequencies higher than 650~GHz show up, where the dispersion in the superconducting niobium's reactance is no longer negligible~\cite{Pronin}.
A slightly larger multimode region could be employed to reduce these tensions.

\subsection{Power coupling efficiency}
The power coupling efficiency in the new design configuration was computed with the model described in~\cite[Section 4]{Cataldo2014}. The ratio of the power emitted by the feed horns to the power received by the antennas is approximately equal to unity. This high efficiency is the result of the absence of any higher-order diffraction peaks in the multimode region due to the relative emitter diameter, $\eta$, being smaller than 1/2~\cite{Hansen2}.

The detailed coupling efficiency of the receiver feeds was not treated. For simplicity, the individual feed structures were mathematically modeled as if the focal surface were subdivided into apertures of equal size.
The current configuration outperforms our previous first-order designs~\cite{Cataldo2014,Cataldo2014b} in terms of efficiency, while simultaneously providing the desired resolving power.


\section{Conclusions}
A design methodology was developed for high-resolution configurations of the $\mu$-Spec multimode region. The design procedure first maximizes the resolving power subject to constraints on geometry, operating frequency range, and performance, thereby determining the order of the spectrometer. This then allows the RMS phase error on the instrument focal plane to be minimized. This work discussed a particular design achieved without constraining the RMS phase error to vanish at preselected points on the focal plane. This led to a configuration with four stigmatic points on the focal plane, a feature which can be used to increase the number of spectrometer channels as the phase error is reduced over a larger angular and spectral bandwidth. The design achieves a maximum RMS phase error equal to 0.1 rad, near-unity coupling, and a resolution of 257 in first order.
Future work will be aimed at employing this design methodology to generate higher-resolution $({\cal R}>500)$ configurations.

\section*{Acknowledgments}
GC would like to thank Prof. Jeffrey A. Hoffman, thesis advisor, for helpful revisions and discussions. The authors gratefully acknowledge the financial support received from the NASA ROSES/APRA program and the Massachusetts Institute of Technology ``Arthur Gelb'' fellowship.

\section*{References}
\label{bib}

\end{document}